\documentclass{Interspeech}

\interspeechcameraready

\title{Spectrotemporal Modulation: Efficient and Interpretable Feature Representation for Classifying Speech, Music, and Environmental Sounds}

\author[affiliation={1}]{Andrew}{Chang}
\author[affiliation={1}]{Yike}{Li}
\author[affiliation={2}]{Iran R.}{Roman}
\author[affiliation={1,3}]{David}{Poeppel}

\affiliation{}{New York University}{USA}
\affiliation{}{Queen Mary University of London}{UK}
\affiliation{}{Max Planck Society}{Germany} 
\email{ac8888@nyu.edu, yikeli@umich.edu, i.roman@qmul.ac.uk, dp101@nyu.edu}

\usepackage{comment}
\usepackage{multirow}

\begin{document}

\maketitle

\begin{abstract}
    
    \noindent Audio DNNs have demonstrated impressive performance on various machine listening tasks; however, most of their representations are computationally costly and uninterpretable, leaving room for optimization. Here, we propose a novel approach centered on spectrotemporal modulation (STM) features, a signal processing method that mimics the neurophysiological representation in the human auditory cortex. The classification performance of our STM-based model, without any pretraining, is comparable to that of pretrained audio DNNs across diverse naturalistic speech, music, and environmental sounds, which are essential categories for both human cognition and machine perception. These results show that STM is an efficient and interpretable feature representation for audio classification, advancing the development of machine listening and unlocking exciting new possibilities for basic understanding of speech and auditory sciences, as well as developing audio BCI and cognitive computing.
\end{abstract}

\section{Introduction}

Sound classification is a fundamental task in machine listening. A critical step in developing powerful audio deep neural networks (DNNs) is converting audio signals into meaningful acoustic feature representations. Common approaches include spectrograms, mel-frequency cepstral coefficients, linear predictive coding, wavelet decomposition, and deep embedded representations (see review \cite{zaman_survey_2023}), and the DNNs based on those feature representations can achieve near-perfect performance on the downstream tasks, including sound classification \cite{gong2021ast,yamnet,vggish}. However, these DNNs usually involve utilizing other pretrained DNNs (mostly computer vision models) and training and inference using millions of parameters, which are computationally costly and demands a high volume of labeled data. Also, while some recent work has proposed lightweight and meaningful DNN representations of audio \cite{richard2025model,roman2024robust, hayes2024review},  most are not easily interpretable nor have any neural, cognitive or auditory basis, which poses challenge when utilizing those models to understand human auditory and speech processing and developing related audio brain-computer interfaces.

\begin{figure*}[htbp]
  \centering
  \centerline{\includegraphics[width=1\textwidth]{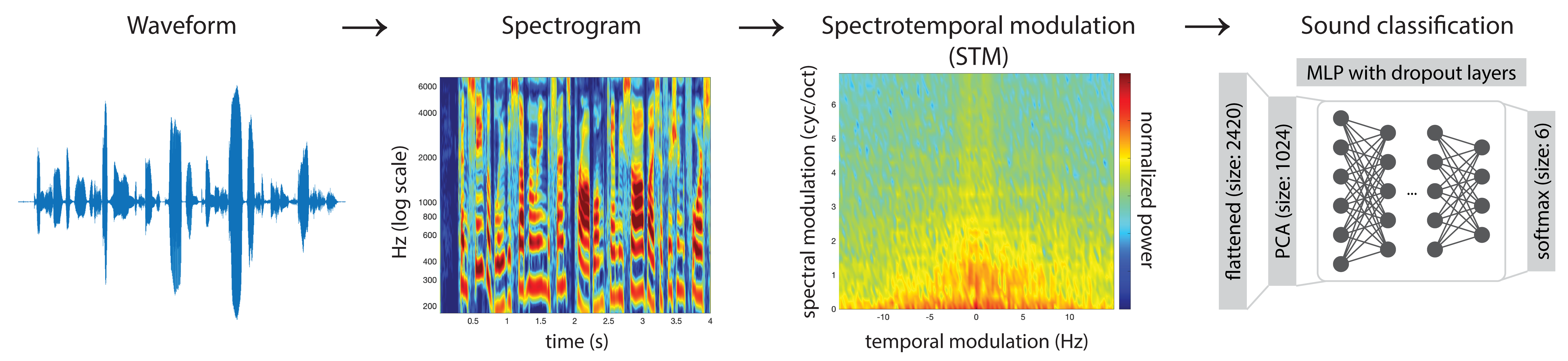}}
  \caption{A processing pipeline for converting an audio signal into a spectrogram, then into STM features for training an MLP classifier.}
  \label{fig:pipe}
\end{figure*}

An acoustic feature that remains underexplored is the spectrotemporal modulation (STM) analysis, which is an analytical signal processing approach that does not require any pretraining. It is a two-dimensional modulation power spectrum that represents how a sound's frequency (spectral modulation) and amplitude (temporal modulation) change over the sound’s spectrogram (Figure \ref{fig:pipe}). The potential application of STM is mostly highlighted by its use in signal processing approaches to enhance and recognize speech signals or predict intelligibility \cite{greenberg1997modulation, edraki2020speech, meyer2011comparing}, classify musical genres \cite{lee2009automatic}, and distinguishing a wide range of sounds (e.g., speech, music, noise) \cite{mesgarani2006discrimination}. These studies showed that STM can effectively represent long (supra-second) complex audio signals. However, the promising potential of STM features to develop data-driven, machine learning solutions (such as DNNs) remains underexplored.

STM is also a fundamental operation carried out by the human auditory cortex in the processing of supra-second complex sounds \cite{albouy_distinct_2020, flinker_spectrotemporal_2019}. Human perceptual studies have shown that different STM feature subspaces represent distinct auditory signals—such as speech, music, and others—for perceptual processing \cite{flinker_spectrotemporal_2019, albouy_distinct_2020, albouy2024spectro, Chang2024}. This aligns with the fact that humans are highly sensitive to these sound categories and robustly represent them in the brain. This suggests the potential of using STM as a neurally and cognitively interpretable feature representation to effectively disentangle different cognitively important sound classes (e.g., speech, music) in machine listening models.

Here we demonstrate the utility of STM features in training a model that exhibits robust and generalizable performance in machine listening tasks, with sound classification performance comparable to other popular audio DNNs. This advances machine listening and speech science research by offering: (1) STM: a neurophysiologically-grounded and cognitively interpretable feature representation, (2) efficient representation of supra-second audio signals across multiple scales of hierarchical spectral and temporal structures, and (3) models using STM as the input representation, eliminating the need for pretraining. 

\section{Approach}

\noindent Dataset and experiments are completely reproducible by using all the code and documentation in the repository.\footnote{\url{https://doi.org/10.5281/zenodo.15521995}}

\subsection{Data preparation}

\noindent The dataset consists of audio tracks featuring speech (tonal and non-tonal), music (vocal and non-vocal), and environmental sounds/others (urban and wildlife). Standard multi-domain audio datasets and benchmarks are not suitable for this study, as most of those sound categories are transient (sub-second), lack long-range temporal structure, and may not necessarily be cognitively important. Therefore, we selected speech, music, and environmental sound categories as they are common in machine listening tasks, essential to human cognition, and human auditory cortex is primarily sensitive to these categories \cite{norman2015distinct}. See the supplementary material in the repository for full details and references.

\subsubsection{Speech corpora}
We included 737,948 single-voice recordings (102,729 and 635,219 of tonal and nontonal language recordings), totaling 1,148.17 hours. Among the 71,768 speakers, 32.83\% were male, 19.99\% were female, and 47.19\% were not reported. The speech recordings included 97 languages or dialects, and a few with different accents or variants, across all continents. The speech corpora included BibleTTS, Buckeye, EUROM1, HiltonMoser2022, LibriSpeech, MediaSpeech, MozillaCommonVoice, Room Reader, Clarity Speech, TAT-Vol2, THCHS-30, TIMIT, TTS-Javanese, and Zeroth-Korean. The categorization of tonal languages follow the definition of complex tonal system of The World Atlas of Language Structures Online. Compared to most previous machine listening models predominantly trained on English, our speech corpora features a significantly greater variety of language samples, as English cannot represent the STM features of all other languages, such as the frequency modulation features of tonal languages.

\subsubsection{Music corpora}
We included 195,067 recordings (2,887.08 hours) played by 22,220 musicians/ensembles. Among the recordings, 128,528 were vocal music and 66,539 were nonvocal music. The music genres include Blues, Classical, Country, Easy Listening, Electronic, Experimental, Folk, Hip-Hop, Infant-Directed Songs, Instrumental, Jazz, Old-Time/Historic, Pop, Rock, Soul, RNB, World (non-Western) and more. The corpora included Albouy2020, Free Music Archive, Garland Encyclopedia of World Music, HiltonMoser2022, IRMAS, ISMIR04, MagnaTagATune, MTG-Jamendo, and NHS2.

To indicate whether each recordings of few music corpora (Free Music Archive, ISMIR04, MagnaTagATune, and MTG-Jamendo) contains vocals or not, as this information was not reported with these corpora, we used \textit{Demucs} \cite{rouard_hybrid_2023}, a hybrid transformer-based music source separation model, to separate the sources and then used YAMNet to identify whether the separated vocal track contains vocals. The Garland corpus was annotated as vocal or not by the first author of this study (an amateur musician), as it is unclear whether \textit{Demucs} can be applied to non-Western music.

\subsubsection{Environmental sound corpora}
The SONYC \cite{mark_cartwright_2020} (n = 10,798) and the Macaulay Library \cite{sullivan2009ebird} (n = 34,565) were included as environmental sound corpora, which featured urban and wildlife sounds, respectively, totaling 468.69 hours from 11,005 sites. Any recordings containing speech, music or voice labels were programmatically removed based on the dataset's original metadata.

\subsection{Audio signal preprocessing and STM analysis}

\noindent We adapted MATLAB (R2023b) STM signal processing pipeline \cite{flinker_spectrotemporal_2019} to extract the STM of all signals (Figure \ref{fig:pipe}). For each four-second excerpt, the sound waveform was downsampled to 16 kHz and transformed into a spectrogram using a filter-Hilbert method, which provides a time-frequency spectrogram estimating the human cochlear processing of a heard sound. Specifically, designed based on cochlear critical bands, the sound waveform was filtered into 128 frequency domain Gaussians, with center frequency \(CF_x\) corresponding to \(CF_x = 440 * 2 ^ {\frac{x-32}{24}}\text{Hz},\) where \(x\) ranged from 1 to 128. They roughly span over 170 -- 7,000 Hz, with the center frequencies of the Gaussian bands logarithmically spaced. The full width at half maximum of each filter band \(BW_x\) corresponded to \(BW_x = 24.7 * (\frac{CF_x * 4.37}{1000} + 1)\). To depict the time-frequency spectrogram, the filtered signals were then Hilbert transformed to extract the analytic amplitude and converted to the dB scale.

\begin{figure*}[htbp]
  \centering
  \centerline{\includegraphics[width=0.92\textwidth]{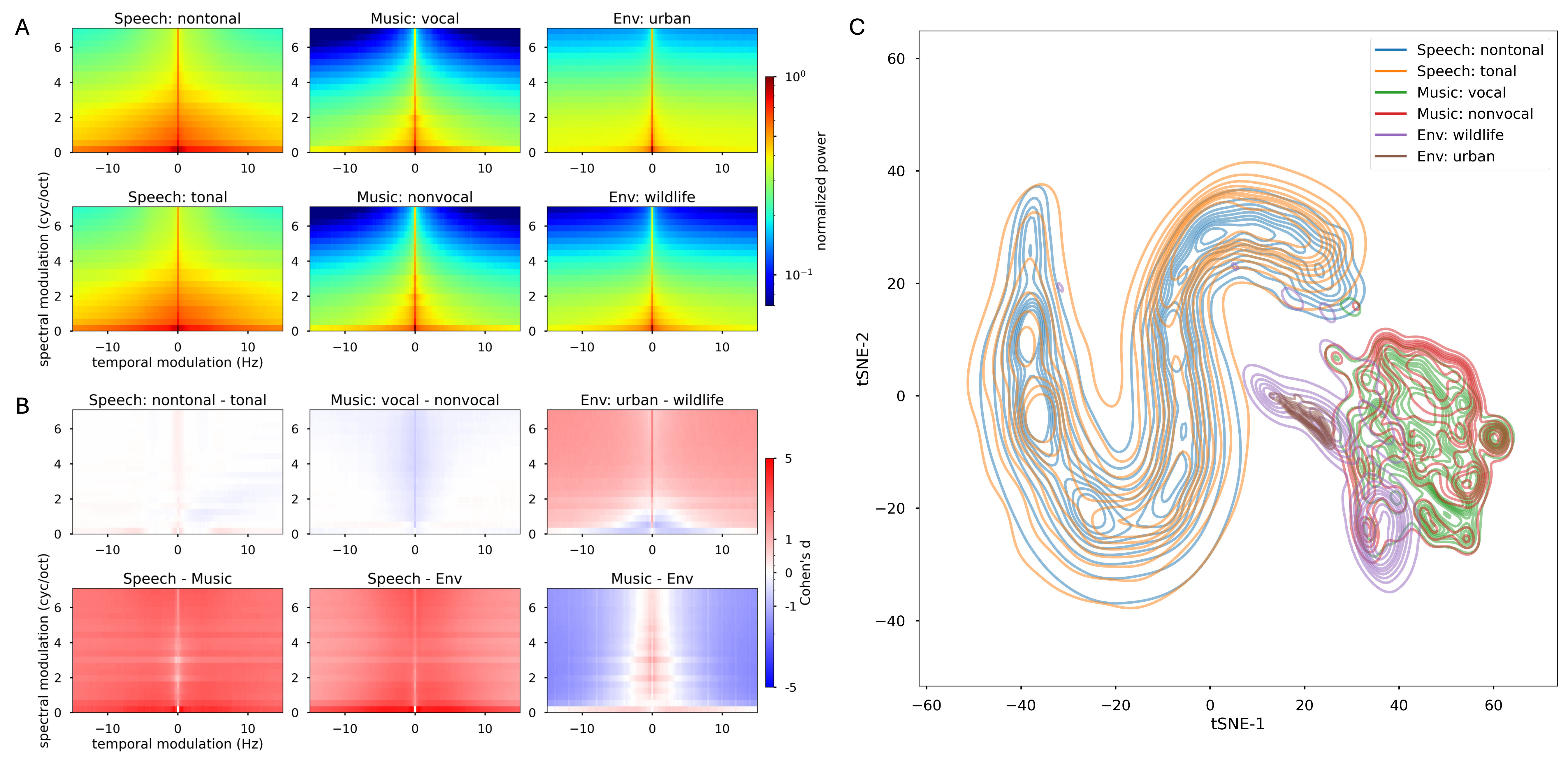}}
  \caption{
  The STM of the different classes in the real-world audio that we study shows clearly separable features. 
  (A) Class-averaged STMs: clear difference between the major classes of speech, music and environmental sounds; (B) Cohen's \textit{d}: the differences between subclasses (tonal vs nontonal speech, vocal vs non-vocal music, and urban vs wildlife environmental sounds) is also captured by STM: speech dominates in higher temporal and lower spectral modulations, music in higher spectral and lower temporal regions, and environmental sounds in the lowest temporal and spectral ranges; (C) A 2-dimensional tSNE projection of the entire high-dimensional dataset we used shows clear separation between the major classes.}
  \label{fig:tsne}
\end{figure*}

The spectrogram matrix was then decomposed into the modulation domain using a two-dimensional fast Fourier transform, which resulted in the STM spectrum. To increase the computational efficiency and reduce the risk of overfitting in the subsequent modeling steps, the two-dimensional STM were cropped and downsampled, resulting in temporal modulation -15 -- 15 Hz (resolution: 0.25 Hz; 121 bins) and spectral modulation 0 -- 7.09 cycles per octave (cyc/oct), resolution: 0.37 cyc/oct; 20 bins, totaling 2420 features. This range was selected to reduce the number of features and avoid overfitting while covering the modulation spectrum of speech and music. 

The audio signals were divided into non-overlapping 4-second chunks for STM analysis and then averaged across chunks, and the chunks shorter than 4 seconds were not analyzed. Chunks containing 1 second or more of silence were excluded from averaging. Up to 120 seconds of signal from each audio sample were used. Last, the dB power of each audio STM was scaled to 0--1. 

\subsection{Multilayer perceptron (MLP) model}
\noindent We chose an MLP neural network architecture because STM features are fixed along the temporal and spectral modulation axes, lacking translational equivalence and temporal patterns. While CNNs and RNNs excel at capturing these aspects in sound classification tasks \cite{zaman_survey_2023}, the fixed nature of STM features makes a simple MLP expressive enough to learn from them.

The data was split into training, validation, and testing sets with a ratio of 8:1:1. Considering the class imbalance, we used \textit{StratifiedGroupKFold()} from \textit{sklearn} to preserve the class ratios while ensuring that the audio recorded from the same speaker, musician, or site were not split into different sets. 

The MLP model of STM was built under \textit{Python} (3.11.9) using \textit{Keras} (3.2.1) with \textit{TensorFlow} (2.15.0) as the backend, and \textit{keras.BayesianOptimization()} was used to implement hyperparameter tuning. For the models trained on STM or melspectrogram features, a PCA was trained on the training set to reduced the size of input dimensions to 1024. While the number of units in the input and output layers was fixed to the number of input features and the number of classes, respectively, the middle layers were subject to hyperparameter tuning: the number of middle layers ranges from 1 to 4. The number of units for each layer ranged from 32 to 512 (step: 32), and L1 regularization ranged \(10^{-12}\)--\(10^{-6}\) (log scale). A dropout layer (rate: 0--0.1) was added after each layer. We used the Adam optimizer, with the learning rate tuned ranged \(10^{-7}\)--\(10^{-4}\) (log scale). The output layer used the softmax activation function, and the rectified linear unit (ReLU) activation function was used in all other layers. Considering the class imbalance, we used categorical focal cross-entropy as the loss function. The tuning objective was maximizing the macro F1 score on the validation dataset. The hyperparameters of the best-performing model were used to retrain with combined training and validation set, and the results on the testing set were reported.

\subsection{Pretrained DNN features for performance comparison}
We trained additional MLP models on AST \cite{gong2021ast} YAMNet \cite{yamnet} and VGGish \cite{vggish} feature embeddings to compare with our STM-based model (identical hyperparameter search spaces for tuning), as their feature representations are widely used in various audio tasks with near-perfect performance in sound classification. The audio samples, resampled to 16 kHz, were fed into each model to obtain their embeddings and then the embeddings were averaged across time frames.  Instead of fine-tuning these models, this approach directly compares their deep embedding representations with STM features. See Table \ref{table:params}.

To determine whether the performance of the STM-based MLP was attributable to the STM features or the MLP architecture, we conducted an additional experiment using melspectrogram features (\textit{librosa}: n\_mels = 32, n\_fft = 2048, hop\_length = 1024). Therefore, comparing MLPs based on STM and melspectrograms helps isolate the contribution of STM.

\begin{table}[tbp]
\caption{Comparison of features we studied by training datapoints and model size. The rightmost column denotes the size of the MLP used on top of the features to do the classification task.}
\begin{center}
\scalebox{0.8}{
\begin{tabular}[c]{ccccc}
    \toprule
    \textbf{Feature}        &  \multicolumn{3}{c}{\textbf{pretrain}}                                                            & \textbf{MLP}  \\
    \cmidrule(lr){2-4}
                            &   datapoints                   & model/architecture                      & params                    & params  \\
    \midrule
    \textbf{STM (full)}     &         0                   & N/A                                     &   0                       &  1.3 M                            \\
    \textbf{STM (reduced)}  &         0                   & N/A                                     &   0                       &  1.1 M                            \\
    AST                     &         2.1 M               & ImageNet ViT                                &   86.2 M                  &  0.7 M                             \\
    YAMNet                  &         2.1 M               & Mobilenet v1                            &   3.8 M                   &  0.8 M                             \\
    VGGish                  &         70 M                & VGG                                     &   72.1 M                  &  0.9 M                             \\
    \bottomrule
\end{tabular}
}
\label{table:params}
\end{center}
\end{table}

\section{Results}

\begin{table*}[htbp]
    \caption{MLP test-set classification performance as a function of using different input features. Note how using simple STM features provides competitive performance compared to deep model representations  (without the need for pretraining while also reducing model size by 7 orders of magnitude). In contrast, using melspectrogram (MelSpec) features does not.}
    \label{Table1}
    \centering
    \scalebox{0.78}{
    \begin{tabular}[c]{c|c|ccccccccc}
        \toprule
        \textbf{Feature}        &\textbf{Undersampling} & \textbf{ROC-AUC}  & \textbf{PR-AUC}       & \multicolumn{7}{c}{\textbf{F1 Score}}\\
        \cmidrule(lr){5-11}
                                &                       &                   &                       & \textbf{macro}   & nontonal speech        & tonal speech      & vocal music       & nonvocal music        & urban env     & wildlife env \\
        \midrule
        \textbf{STM}            &No                     &  .988             &  .937                 &  .808            & .938                   & .676              & .814              & .638                  & .902          & .880\\
        AST                     &No                     &  .995             &  .975                 &  .884            & .970                   & .816              & .835              & .702                  & .993          & .984\\ 
        YAMNet                  &No                     &  .990             &  .952                 &  .849            & .948                   & .690              & .848              & .737                  & .907          & .965\\ 
        VGGish                  &No                     &  .993             &  .966                 &  .871            & .958                   & .759              & .856              & .741                  & .937          & .972\\ 
        \textit{MelSpec}        &No                     &  .944             &  .740                 &  .479            & .878                   & .300              & .700              & .346                  & .000          & .648\\ 
        \midrule
        \textbf{STM}            &Yes                    &  .973             &  .869                 &  .807            & .790                   & .818              & .816              & .635                  & .894          & .889\\
        AST                     &Yes                    &  .986             &  .930                 &  .875            & .862                   & .882              & .842              & .709                  & .978          & .977\\ 
        YAMNet                  &Yes                    &  .980             &  .905                 &  .835            & .795                   & .819              & .860              & .739                  & .833          & .965\\ 
        VGGish                  &Yes                    &  .984             &  .923                 &  .862            & .822                   & .842              & .863              & .749                  & .923          & .972\\ 
        \textit{MelSpec}        &Yes                    &  .900             &  .588                 &  .530            & .568                   & .692              & .725              & .445                  & .000          & .749\\ 
        \bottomrule
    \end{tabular}%
    }
\end{table*}

\subsection{STM feature representation and statistical analyses}
\noindent The class-averaged STMs and the distribution of individual STMs in tSNE space (Figure \ref{fig:tsne}) showed that, even without supervised neural network modeling, the STM feature representation can effectively distinguish sound categories in the tSNE space. Specifically, the speech audio was more distinguishable from the others, as confirmed by the univariate statistical effect size, Cohen's \textit{d}. This aligns with the widely held view that speech is a unique audio signal processed distinctly by the brain \cite{hickok2007cortical}. Additionally, the low-temporal and low-spectral modulation ranges most saliently reflected the class differences. 

\subsection{Model performances across feature representations}
\noindent The classification performances are summarized in Table~\ref{Table1}. The results showed that the best model based on STM features achieved a ROC-AUC of 0.988 (which is considered a more appropriate and sophisticated metric for evaluating imbalanced datasets \cite{jeni_facing_2013, johnson2019survey}), and macro-F1 score of 0.808, which were much higher than those of melspectrogram and comparable to other audio DNNs. These suggest that STM is an efficient and effective feature representation. 

To further examine whether class imbalance, caused by an excess of nontonal speech samples, would bias the results, we undersampled the nontonal speech samples to 100k. Although this slightly reduced most metrics across the models, it significantly balanced the F1 scores of the STM model across categories, performing markedly better than the melspectrogram-based model. This shows that the STM features faithfully represents each of these sound categories.

\subsection{Ablation study}

\begin{figure}
  \centering
  \centerline{\includegraphics[width=0.5\textwidth]{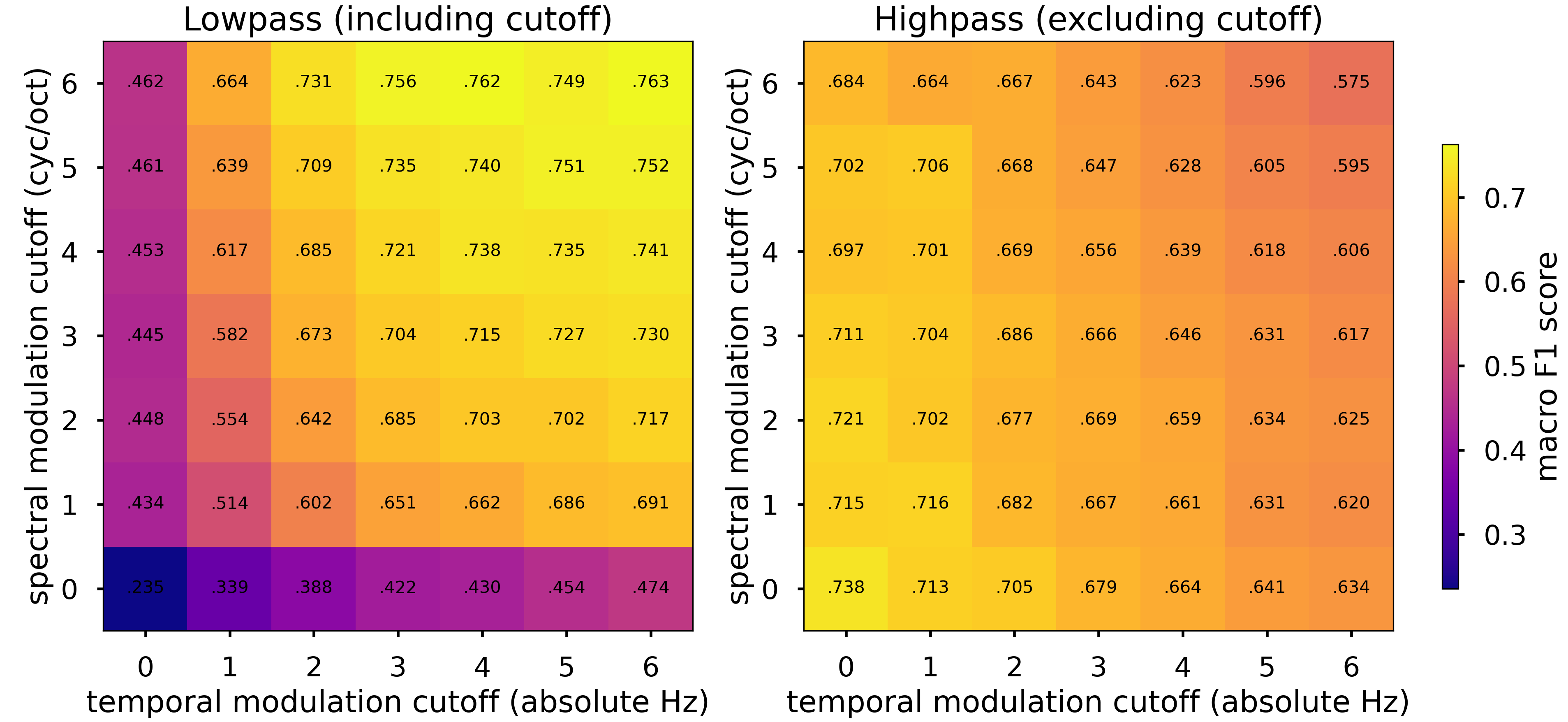}}
  \caption{Macro-F1 score on the test-set under different lowpass (left) and highpass (right) STM ablations. The lowpass plot indicates that the model trained on STM $\leq$$\pm4$ Hz and $\leq 6$ cyc/oct (the "reduced" model) performed comparably to the model trained on the full STM space, indicating the importance of lower spectral and temporal modulation features for sound classification. Note that the plots were truncated at 6 cyc/oct and $\pm$6 Hz for visualization purposes, but the model still included STM features beyond these limits.}
  \label{fig:ablation}
\end{figure}

\noindent We conducted an ablation study to identify which STM subregion is critical for model performance by training additional models (no undersampling) using the same settings (Figure \ref{fig:ablation}). The lowpass plot shows that the model trained on STM $\leq \pm 4$ Hz and $\leq 6$ cyc/oct (the "reduced" model) performed comparably to the model trained on the full STM feature space, suggesting that this defines the model's critical STM subspace. The highpass plot also confirms the importance of lower spectral and temporal modulations for sound classification, as the model's performance drops significantly even when only a small portion of the STM features is excluded. These results align with Cohen's \textit{d} (Figure \ref{fig:tsne}B), which shows the lower modulations regions have the strongest statistical difference among classes.

\section{Discussion and Conclusion}

\noindent We demonstrate the potential of STM features for building powerful, efficient, and interpretable audio model to classify speech, music, and environmental sounds. Our STM-based neural network model, without any pretraining, achieved exceptional performance, comparable to the models trained on the deep embedded features of audio DNN models, which required pretraining on millions of parameters with millions of labeled samples (Table \ref{table:params}). This highlights the efficiency of STM features in representing complex supra-second audio signals, with even greater potential to be unlocked through larger datasets.

STM feature representations have the following advantages: (1) STM has proven effective in capturing complex supra-second audio signals, such as speech and music \cite{greenberg1997modulation, edraki2020speech, meyer2011comparing, lee2009automatic, mesgarani2006discrimination}, unlike most of the audio DNNs, which focus on representing sub-second features. (2) It intrinsically supports variable-length audio signal, and the spectrotemporal resolution can be adjusted for any specific tasks. (3) STM represents audio signals with hierarchical spectral and temporal structures across multiple scales, such as sentences, speech prosody, or musical phrases. In contrast, using RNNs, LSTMs, or transformers for such tasks can be computationally more expensive due to their higher model complexity. However, since STM analyzes the entire audio segment, other audio DNN embeddings may be more effective for representing transient sound events.

Finally, this work highlights the potential of building a DNN with neurophysiologically interpretable feature representations, offering a more direct link between machine perception and human brain and cognition, compared to other audio DNNs \cite{tuckute_many_2023}. This work paves the way for promising advancements in neuromorphic and cognitive computing, audio-related brain-computer applications, including sound and speech reconstruction \cite{santoro2017reconstructing, martin2014decoding}.

\section{Acknowledgements}
A.C. is supported by National Research Service Award, NIDCD/NIH (F32DC018205), Leon Levy Scholarships in Neuroscience, Leon Levy Foundation and New York Academy of Sciences, and GRAMMY Museum Grants Program. The funders have no role in study design, data collection and analysis, decision to publish, or preparation of the manuscript. This work was supported in part through the NYU IT High Performance Computing resources, services, and staff expertise. We thank Adeen Flinker, Brian McFee, and Patrick Savage for their assistance, as well as the researchers and volunteers for their contributions to the audio corpora.

\bibliographystyle{IEEEtran}
\bibliography{mybib}

\end{document}



\begin{table}[ht]
    \caption{Speech corpora}
    \label{Table_speech}
    \centering
    {
    \resizebox{1\textwidth}{!}{%
    \begin{tabular}{lrm{7.5cm}l} \toprule
        Name                                            &\# of samples  &Languages                                                  & License\\\midrule
        BibleTTS \cite{meyer_bibletts_2022}             &113,348        &Akuapem, Asante, Ewe, Hausa, Lingala, Yoruba               & CC BY-SA\\
        Buckeye \cite{pitt_buckeye_2005}                &255            &English-US                                                 & Noncommercial uses\\
        EUROM1 \cite{chan_eurom-_1995}                  &6,623          &Danish, Dutch, English, French, German, Norwegian, Swedish & Research uses \\
        HiltonMoser2022 \cite{hilton_acoustic_2022}     &785            &English                                                    & CC BY-NC-SA 4.0\\
        LibriSpeech \cite{panayotov_librispeech_2015}   &27,144         &English                                                    & CC BY 4.0\\
        MediaSpeech \cite{kolobov_mediaspeech_2021}     &10,022         &Arabic, French, Spanish, Turkish                           & CC BY 4.0 \\
        MozillaCommonVoice (16.1)\footnotemark[1] \cite{ardila_common_2020} &491,982 &Abkhaz, Arabic, Armenian, Bashkir, Basque, Belarusian, Bengali, Breton, Bulgarian, Cantonese, Catalan, Central Kurdish, Chuvash, Czech, Danish, Dhivehi, Dutch, English, Esperanto, Estonian, Finnish, French, Frisian, Galician, Georgian, German, Greek, Guarani, Hakha Chin, Hindi, Hungarian, Igbo, Indonesian, Irish, Italian, Japanese, Kabyle, Kazakh, Kinyarwanda, Kurmanji Kurdish, Kyrgyz, Latgalian, Latvian, Lithuanian, Luganda, Malayalam, Maltese, Mandarin-China, Mandarin-Taiwan, Meadow Mari, Mongolian, Occitan, Odia, Persian, Polish, Portuguese, Romanian, Russian, Serbian, Spanish, Swahili, Swedish, Taiwanese, Tamil, Tatar, Thai, Turkish, Ukrainian, Urdu, Uyghur, Uzbek, Vietnamese, Welsh, Yoruba & CC BY-SA 4.0 \\
        Primewords Chinese Corpus Set 1 \cite{primewords_information_technology_co_primewords_2018} &47,166 &Mandarin-China         & CC BY-NC-ND 4.0 \\
        Room Reader \cite{reverdy_roomreader_2022}      &146            &English                                                    & Noncommercial uses \\
        Clarity Speech \cite{graetzer_dataset_2022}     &555            &English-UK                                                 & CC BY-SA 4.0\\
        TAT-Vol2\footnotemark[2] \cite{liao_formosa_2020} &793          &Taiwanese                                                  & Noncommercial uses \\
        THCHS-30 \cite{wang_thchs-30_2015}              &13,327         &Mandarin-China                                             & Apache License v.2.0\\
        TIMIT \cite{garofolo_john_s_timit_1993}         &479            &English-US                                                 & Research uses\\
        TTS-Javanese \cite{sodimana_step-by-step_2018}  &3,068          &Javanese                                                   & CC BY-SA 4.0 \\
        Zeroth-Korean \cite{jo_korean_nodate}           &22,255         &Korean                                                     & CC BY 4.0\\
        \bottomrule
    \end{tabular}%
    }%
    }
\end{table}
\footnotetext[1]{Including a language if the number of voices is greater than 1000. Only verified samples were included. If the number of files is greater than 10,000, randomly sample 10,000 files.}
\footnotetext[2]{Only the free-access samples were included.}


\begin{table}
    \caption{Music corpora}
    \label{Table_music}
    \centering
    \resizebox{\textwidth}{!}{%
    \begin{tabular}[c]{lrl}\toprule
        Name                                                                &\# of samples  & License \\\midrule
        Albouy2020\cite{albouy_distinct_2020}                               &123            & Noncommercial uses\\
        FMA (large)\footnotemark[3] \cite{defferrard_fma_2017}              &106,257        & CC BY 4.0\\
        The Garland Encyclopedia of World Music \cite{stone_garland_2017}   &296            & Fair use\\
        HiltonMoser2022 \cite{hilton_acoustic_2022}                         &803            & CC BY-NC-SA 4.0\\
        IRMAS \cite{bosch_comparison_2012}                                  &2,838          & CC BY-NC-SA 3.0\\
        ISMIR04 \cite{cano_vila_ismir_2006}                                 &2,185          & CC BY-NC-ND 2.5\\
        MagnaTagATune \cite{law_evaluation_2009}                            &25,859         & CC BY-NC-SA 3.0\\
        MTG-Jamendo \cite{bogdanov_mtg-jamendo_2019}                        &55,699         & CC BY 3.0 ES\\
        NHS2 \cite{bertolo_cross-cultural_2023}                             &1,007          & CC BY-NC-SA 4.0\\
        \bottomrule
    \end{tabular}%
    } 
\end{table}
\footnotetext[3]{This set only includes the first 30 seconds of each audio.}

\clearpage 
\bibliographystyle{IEEEtran}
\bibliography{mybib}